\documentclass[12pt]{article}
\usepackage[english,russian]{babel}
\usepackage{latexsym}
\usepackage{amsthm}
\usepackage{graphicx}
\usepackage{subeqnarray}
\usepackage{amsmath}
\usepackage{amsfonts}
\usepackage{amscd}
\usepackage{amsthm}
\usepackage{amssymb}
\usepackage{latexsym}
\newcommand{\nc}{\newcommand*}

\nc{\reff}[1]{(\ref{#1})}
\nc{\ds}{\displaystyle}
\nc{\ts}{\textstyle}
\nc{\nn}{\nonumber}

\usepackage[cp1251]{inputenc}
\inputencoding{cp1251}
\usepackage[T2B]{fontenc}
\usepackage[russian,english]{babel} 
\begin{document}
\selectlanguage{english}

{\large\bf V.V. Borzov
\footnote{Department of Mathematics, St.Petersburg State University of Telecommunications, 22-1, Prospekt
Bolshevikov, St.Petersburg, 193232, Russia,\hfill\break  e-mail: borzov.vadim@yandex.ru}
E.V. Damaskinsky \footnote{Mathematical Department, VI(IT),  Zacharievskaya 22,
191123 Russia,\hfill\break e-mail: evd@pdmi.ras.ru}}
\bigskip

\begin{flushright} To our teacher \qquad\quad\quad

Vasily Mikhailovich Babich

with deep respect \qquad\qquad\quad
\end{flushright}
\bigskip

\begin{center}
{\Large\bf Calculation the Mandel parameter
\\[0.2 cm]
for an oscillator-like system generated by
\\[0.2 cm]
generalized Chebyshev polynomials}
\end{center}
\bigskip

\begin{quote}
In this paper, we calculate the Mandel parameter $Q_M$
for an oscillator-like system generated by generalized Chebyshev polynomials \cite{01}, \cite{02}, \cite{03}.
The sign of the Mandel parameter $Q_M$ characterizes the deviation of the excitation
statistics from the Poisson one. This work is a continuation of our works \cite{04}, \cite{05}.
\end{quote}
\bigskip

\vspace{1cm}

\section{Introduction}
The Mandel parameter was introduced in \cite{06}. For a standard harmonic oscillator in
Fock space $\mathcal{H}_F$ it is calculated by the formula
\begin{equation}\label{01}
Q=\frac{\langle (\Delta(n))^2\rangle -\langle n\rangle}{\langle n\rangle},
\end{equation}
where $\Delta(n)=\sqrt{\langle n^2\rangle -\langle n\rangle^2}$, and $n=a^+a$ is the operator of the
number of particles (excitations). The sign of the Mandel parameter determines the nature of the
deviation of the excitation statistics from the Poisson one.
For coherent states, $Q=0$ (Poisson statistics), $Q>0$ for ordinary (classical) states
(super-Poisson case) and $Q<0$ for non-classical States (sub-Poisson case).
When $Q<0$, the phenomenon of anti-bunching occurs
$\langle (\Delta(n))^2\rangle <\langle n\rangle$. For physical aspect of bunching and anti-bunching
of photons in quantum optics, see \cite{07}. Detailed overview
non-classical states in quantum optics, is given in \cite{08}. Where a large class of non-classical states:
coherent and squeezed states, states with minimal uncertainty, intelligent states,
binomial states, deformed coherent states, etc. are  considered.

In \cite{05}, we showed that for coherent states of oscillator-like systems,
generated by known classes of orthogonal polynomials, the Mandel parameter can take
both positive and negative values. In particular, for oscillators generated by
Charlier polynomials \cite{09}, as well as for the standard oscillator (Hermite polynomials)
the Mandel parameter takes a null value ($Q_M=0;$ Poisson statistics).
For Laguerre and Meixner polynomials $Q_M<0$ (sub-Poisson statistics).
For Chebyshev and Legendre polynomials $Q_M>0$ (super-Poisson statistics).
In the case of Gegenbauer and Kravchuk polynomials, the sign of the Mandel parameter depends on
the size of the eigenvalue of the annihilation operator corresponding to the
coherent state. In the case of deformed $q$-Hermite polynomials
the sign $Q_M$ is determined by the value of the deformation parameter $q.$ Namely,
$Q_M<0$ for $0<q<1,$ a for $q>1$  the Mandel parameter is positive ($Q_M>0$).

Note that in \cite{05} we used Barut - Girardello coherent states \cite{10},
which are defined as eigenstates of annihilation operator (in the case of coherent states
for groups - as eigenstates of destruction operator).
These states can also be defined for generalized oscillators associated with
classical orthogonal polynomials and their $q$-analogues, in the case when the corresponding
the Hilbert space is infinite-dimensional. Similar results are valid also for coherent states of
the Klauder - Gaseau type \cite{11}.

In this paper we show on example of a generalized Chebyshev oscillator \cite{01} that even a
one-dimensional perturbation of the Jacobi matrix for Chebyshev polynomials can change the
statistics of coherent states (from super-Poisson to sub-Poisson).
For the simplest case of generalized Chebyshev polynomials $Ch_{n}(z;1;a)$ we prove that 
critical value of the perturbation parameter $a=\frac{1}{\sqrt{2}}.$
For values of $a<\frac{1}{\sqrt{2}}$, the statistics can become sub-Poisson.

\section{Generalized Chebyshev polynomials}
Generalized Chebyshev polynomials $Ch_{n}(z;k;a),\, k\geq 1,$ are defined by the recurrence relations:
\begin{multline}\label{01a}
b_{n}Ch_{n+1}(z;k;a)+b_{n-1}Ch_{n-1}(z;k;a)=zCh_{n}(z;k;a),\quad n\geq 0,\\
Ch_{0}(z;k;a)=1,\quad Ch_{-1}(z;k;a)=0,\qquad\qquad\qquad\qquad
\end{multline}
where $b_{n}=1,$ for $n\neq {k-1}$  and $b_{k-1}=a$.
Using the expression obtained in \cite{05} for polynomials related to
with the relation \reff{01a} (and with the corresponding Jacobi matrix), we have
\begin{equation*}
Ch_{n}(z;k;a)=
\!\!\!\sum_{m=0}^{Ent(\frac{n}{2})}\frac{(-1)^m}{\sqrt{[n]!}}b_{0}^{2m-n}\beta_{2m-1,n-1}z^{n-m},
\end{equation*}
where $\beta_{-1,n-1}=1,\, n\geq 0,$ and
\begin{equation*}
\beta_{2m-1,n-1}=\sum_{k_1=2m-1}^{n-1}\!\!\![k_1]!\sum_{k_2=2m-3}^{k_{1}-2}[k_2]!\cdots
\sum_{k_m=1}^{k_{m-1}-2}[k_m]!
\end{equation*}
for all $m\geq 1.$ Here $[s]=\ds\frac{b_{s-1}^2}{b_{0}^2},$ and $Ent(x)$ --- the integer part of $x$.

As an example, we give the last formulas in the case of $k=1$, (denoting $\Psi_{n}(z)=Ch_{n}(z;1;a)$):
\begin{multline*}
\qquad\qquad\qquad\qquad\qquad\Psi_{0}(z)=1,\quad \Psi_{1}(z)=\frac{z}{a},\\
\Psi_{n}(z)=\frac{z^n}{a}-\frac{n+(a^{2}-2)}{a}z^{n-2}+\qquad\qquad\qquad\qquad\qquad\qquad\\
\sum_{m=2}^{Ent(\frac{n}{2})}(-1)^{m}\frac{(n-m-1)!(n+m(a^{2}-2))}{(n-2m)!m!a}z^{n-2m},\quad n\geq 2.
\end{multline*}

Jacobi matrix $J_k$ associated with generalized Chebyshev polynomials\hfill\break $Ch_{n}(z;k;a)$ has the following form.
All its elements $j^k_{i, j}$ are equal to zero except for the elements on the first over-diagonal and
the first under-diagonal, which are equal to
\begin{equation*}
\left\{\begin{aligned}
j^k_{i,i+1}=j^k_{i+1,i}=&1,\quad i\neq k;\\
j^k_{k,k+1}=j^k_{k+1,k}=&a,\quad i= k,
\end{aligned}\right.
\end{equation*}
In other words, the parameter $a$ stands in the over-diagonal and under-diagonal at the $k$-th place from the top, and all
other elements are equal to $1$. As an example, we give $J_k$ for $k=1$ and $k=4$
\begin{equation*}
J_1=\begin{bmatrix}
0&a&0&0&0&0&\cdots&\cdots\\
a&0&1&0&0&0&\ddots&\cdots\\
0&1&0&1&0&0&\ddots&\cdots\\
0&0&1&0&1&0&\ddots&\cdots\\
0&0&0&1&0&1&\ddots&\cdots\\
\hdotsfor{8}
\end{bmatrix}
\qquad
J_4=\begin{bmatrix}
0&1&0&0&0&0&\cdots&\cdots\\
1&0&1&0&0&0&\ddots&\cdots\\
0&1&0&1&0&0&\ddots&\cdots\\
0&0&1&0&a&0&\ddots&\cdots\\
0&0&0&a&0&1&\ddots&\cdots\\
\hdotsfor{8}
\end{bmatrix} .
\end{equation*}

In this paper, we restrict ourselves to the cases of $k=1$ and $k=2$.

\section{Generalized Chebyshev oscillator \\ (in the case of $\mathbf{k=1}$)}
Let $a>0$, $\mathcal{H}_a=L^2(\mathbb{R};\mu_{a})$ --- a fixed Hilbert space, and
$\lbrace\varphi_n(x)\rbrace_{n=0}^\infty$
be a system of polynomials orthonormal with respect to the measure $\mu_a,$ where
\begin{equation*}
d\mu_{a}(x)=\frac{1}{2\pi}
\left\{\begin{aligned} & \frac{a^2\sqrt{4-x^2}}{a^4-(a^2-1)x^2}dx,\quad\text{if}\quad |x|\leq 2,\\
&\quad  0,\quad\quad\quad\quad\quad\quad\quad\,\,\text{if}\quad |x|>2.\end{aligned}\right.
\end{equation*}
Then, as follows from \cite{02} (see also \cite{03}), the polynomials $\varphi_n(x)$ are
generalized Chebyshev polynomials $\Psi_{n}(x)==Ch_{n}(x;1;a)$ (for the case $k=1$) and recurrent 
relations \reff{01a} take the form:
\begin{gather*}
a\Psi_{1}(x)=x\Psi_{0}(x),\quad \Psi_{2}(x)+a\Psi_{0}(x)=x\Psi_{1}(x),\nn\\
\Psi_{n+1}(x)+\Psi_{n-1}(x)=x\Psi_{n}(x),\quad n\geq 2,\\
\Psi_{0}(x)=1,\quad \Psi_{-1}(x)=0,\nn
\end{gather*}

The first few polynomials are equal
\begin{align*}
\Psi_0=1;&\qquad \Psi_3=\frac{x}{a}\left(x^2-(a+1)\right);\\
\Psi_1=\frac{x}{a};&\qquad \Psi_4=\frac{x^4-(2+a)x^2+a}{a};\\
\Psi_2=\frac{x^2-a}{a};&\qquad \Psi_5=\frac{x}{a}\left(x^4-(3+a)x^2+(1+2a)\right);
\end{align*}
\begin{align*}
\Psi_6=&\frac{x^6-(4+a)x^4+3(a+1)x^2-a}{a};\\
\Psi_7=&\frac{x}{a}\left(x^6-(5+a)x^4+(6+4a)x^2-(1+3a)\right);\\
\Psi_8=&\frac{x^8-(6+a)x^6+(10+5a)x^4-(4+6a)x^2+a}{a};\\
\Psi_9=&\frac{x}{a}\left(x^8-(a+7)x^6+(15+6a)x^4-10(1+a)x^2+(1+4a)\right);\\
\Psi_{10}=&\frac{x^{10}-(8+a)x^8+(21+7a)x^6-(20+15a)x^4+(5+10a)x^2-a}{a}.
\end{align*}

In \cite{12}, a method for constructing an oscillator-like algebra
$\mathfrak{A}_{\Psi}$ corresponding to this system of polynomials was proposed. Polynomials
$\lbrace\Psi_n(x)\rbrace$ $0\leq n<\infty$
form the Fock basis for this algebra, $\mathfrak{A}_{\Psi}$ in the Fock space $\mathcal{H}_a$.
Generators $a_{\mu_a}^{+},\, a_{\mu_a}^{-},\, N_{\Psi}$ of algebra
$\mathcal{A}_\Psi$ in this Fock representation acts as follows
\begin{equation*}
a_{\mu_a}^{+}\Psi_n=\sqrt{2}b_n\Psi_{n+1},\quad a_{\mu_a}^{-}\Psi_n=\sqrt{2}b_{n-1}\Psi_{n-1},
\quad N_{\Psi}\Psi_n=n\Psi_n,
\end{equation*}
where
\begin{equation*}
b_{-1}=0,\quad b_0=a,\quad b_n=1,\quad  n\geq1.
\end{equation*}

Let $I$ --- be a unit operator in the Hilbert space $\mathcal{H}_a.$
Define $B_{\Psi}(N_{\Psi})$ as an operator-valued function defined by the equalities
\begin{equation*}
B_{\Psi}(N_{\Psi})\Psi_{n}=b_{n-1}^{2}\Psi_{n},\quad B_{\Psi}(N_{\Psi}+I)\Psi_{n}=b_{n}^{2}\Psi_{n},
\quad n\geq 0.
\end{equation*}
Then the algebra of the generalized Chebyshev oscillator $\mathfrak{A}_{\Psi}$ is
generated by the operators $a_{\Psi}^{\pm},$ $N_{\Psi}$ and $I$ satisfying the relations
\begin{gather*}
a_{\mu_a}^{-}a_{\mu_a}^{+}\Psi_{n}=2B_{\Psi}(N_{\Psi}+I),\quad
a_{\mu_a}^{+}a_{\mu_a}^{-}\Psi_{n}=2B_{\Psi}(N_{\Psi}),\quad\\
[N_{\Psi},a_{\mu_a}^{\pm}]=\pm a_{\mu_a}^{\pm},
\end{gather*}
and by the commutators of these operators.

Similarly, we can define the algebra of the generalized Chebyshev oscillator $\mathfrak{A}_k$ corresponding to
generalized Chebyshev polynomials $Ch_{n}(z;k;a)$ for an arbitrary integer $ k\geq 1.$
Clearly, $\mathfrak{A}_{\Psi}=\mathfrak{A}_1.$

For the algebra $\mathfrak{A}_k$ we define coherent states
of the Barut-Girardello type as usual \cite{10} (see also \cite{05})
\begin{gather*}
a_{k}^{-}|z\rangle=z|z\rangle,\quad\\
|z\rangle=N^{-\frac{1}{2}}(|z|^2)\sum_{n=0}^{\infty}
\frac{z^n}{(\sqrt{2}b_{n-1})!}Ch_{n}(z;k;a),
\end{gather*}
where the normalizing factor has the form
\begin{equation}\label{02}
N(|z|^2)=\sum_{n=0}^{\infty}
\frac{|z|^{2n}}{(2b_{n-1}^2)!}.
\end{equation}

For computing the Mandel parameter in coherent states $|z\rangle$
for in \cite{05}, the following formula was obtained
\begin{equation}\label{03}
Q_M(x)=x\left(\frac{N^{''}(x)}{N^{'}(x)}-\frac{N^{'}(x)}{N(x)}\right), \quad
(|z|^2=x).
\end{equation}
Using the values of the coefficients $b_n$ of recurrent relations
\reff{01a}  for generalized Chebyshev polynomials
$Ch_{n}(z;k;a),\, k\geq 1,$ and the formulas \reff{02} and \reff{03},
we calculate the Mandel parameter $Q_M(x;k;a)$
for coherent states of an algebra $\mathfrak{A}_k$ of the generalized Chebyshev oscillator.

A further task is to determine the sign-constant regions
of the Mandel parameter $Q_M(x; k; a)$ depending on the values of
the perturbation parameter $a$ of the Jacobi matrix for generalized Chebyshev polynomials.

\section{Computation of the Mandel parameter \\ (in the case of $\mathbf{k=1}$)}
We want to calculate the Mandel parameter $Q_M(x;1;a)$ for coherent states of the algebra
$\mathfrak{A}_1$ of the generalized Chebyshev oscillator.
Let's start by calculating the normalizing factor $N_1.$
From the formula \reff{02} it follows that
\begin{equation}\label{04}
N_1(x)=\sum_{n=0}^{\infty}\frac{x^{n}}{(2b_{n-1}^2)!},
\end{equation}
where $b_{-1}=0,\, b_0=a,\, b_n=1 , n\geq 1.$ Expression $(2b_{n-1}^2)!$ 
is understood as a factorial by index,
that is $(2b_{n-1}^2)!=(2b_0)(2b_1)...(2b_{n-1})$ and $(2b_{-1})!=0!=1$
because $b_{-1}=0.$ Then we get
\begin{equation}\label{05}
N_1(x)=a^{-2}\left(a^2 -1 +\frac{2}{2-x}\right), \quad 0<x<2.
\end{equation}

The restriction $0<x<2$ is due to the convergence region of the power series in \reff{04}.
Differentiating \reff{05}, we find
\begin{equation*}
N_1^{\prime}(x)=2a^{-2}(2-x)^{-2},\quad N_1^{''}(x)=4a^{-2}(2-x)^{-3}.
\end{equation*}
From \reff{03},\reff{04} and \reff{05} we get the desired formula
\begin{equation}\label{06}
Q_M(x;1;a)=\frac{2x}{2-x}\left(1- \frac{1}{x+a^2(2-x)}\right),\quad 0<x<2.
\end{equation}
Let us study the sign of the Mandel parameter for different values of the perturbation parameter $a.$
Note that for $0<x<2$, the multiplier before the bracket in the right hand part of
the equality \reff{06} is positive. Therefore,
\begin{equation}\label{07}
\text{sign}(Q_M(x;1;a))=\text{sign}\left(1- \frac{1}{x+a^2(2-x)}\right), \quad 0<x<2.
\end{equation}
In addition, for $0<x<2$, the denominator of the fraction,
standing on the right hand side of the equality \reff{07} is also positive. Therefore,
\begin{equation}\label{08}
\text{sign}(Q_M(x;1;a))=\text{sign}\left(x(1-a^2)+2a^2-1\right), \quad 0<x<2.
\end{equation}

Let us show that $Q_M(x;1;a)>0$ for $a^2\geq\frac{1}{2}$\, for all $x\in (0;2)$, and when
$a^2<\frac{1}{2}$ if $x>\ds\frac{1-2a^2}{1-a^2}$. For $\frac{1}{2}\leq a^2\leq 1$ \, ($0<x<2$)
it is obvious that the inequality
\begin{equation}\label{09}
x(1-a^2)>1-2a^2.
\end{equation}
hold true. For $a^2>1$, we rewrite the inequality \reff{09} as
equivalent inequality
\begin{equation*}
2a^2-1>x(a^2-1),
\end{equation*}
which in turn is equivalent to the inequality
\begin{equation*}
x<\frac{2a^2-1}{a^2-1}=2+ \frac{1}{a^2-1}.
\end{equation*}
The last inequality obviously holds for $a^2\geq 1$
and $0<x<2$. Hence, the fairness of the inequality
$Q_M(x;1;a)>0$ for $a^2\geq \frac{1}{2}$\, ($0<x<2$) is proved.

If $a^2<\frac{1}{2}$, then it follows from the inequality \reff{09} that
$Q_M(x;1;a)>0$, when
\begin{equation*}
x>\frac{1-2a^2}{1-a^2}.
\end{equation*}
Hence, $Q_M(x;1;a)<0$ for $a^2<\frac{1}{2}$ and
\begin{equation*}
0<x<\frac{1-2a^2}{1-a^2}.
\end{equation*}

As an illustration, we present the graphs of $Q_M(x;1;a)$ at $a=0.5$ and $a=0.65$ (on the left) and at
$a=1$ and $a=2$ (on the right).

\begin{figure}[h]
\centering
\hspace{-0.5cm}
\includegraphics[width=5.8cm]{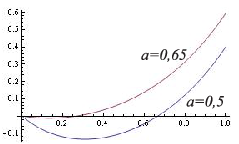}
\hspace{0.5cm}
\includegraphics[width=5.6cm]{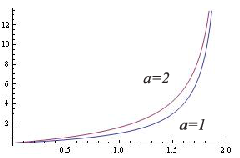}
\caption{Plots of the functions $Q_M(x;1;a)$}
\end{figure}

The left graph shows that for $a<\frac{1}{\sqrt{2}}$, when $a$ decreases, the area with
the negative value of the Mandel parameteriz increases, and for $a>
\frac{1}{\sqrt{2}}$, there is no such area.

\section{Calculating the Mandel parameter\\ for $\mathbf{k\geq\,1}$}
For further consideration, it is convenient to slightly transform the formula \reff{03}.
Let's start by calculating the normalizing factor $N_k(x;a^2)$. According to the formula \reff{02} we have
\begin{equation*}
N_k(x;a^2)=\sum_{n=0}^{\infty}\frac{x^n}{(2b_{n-1}^{\,\,2})!}\qquad k\geq1,
\end{equation*}
where $x=|z|^2>0$, $b_{k-1}=a>0$, $b_n=1$ for $n\neq (k-1)$ and $b_{-1}=0$. The convergence condition of an
infinite geometric progression leads to the restriction $0<x<2$. Under this condition, the expression
$N_k(x;a^2)$ can be rewritten as
\begin{multline*}
\qquad\qquad N_k(x;a^2)=\sum_{n=0}^{k-1}\frac{x^n}{2^n}+\frac{1}{a^2}\sum_{n=k}^{\infty}\frac{x^n}{2^n}=\\
\frac{1}{a^2}\,\frac{1}{1-\frac{x}{2}}+
\left(1-\frac{1}{a^2}\right)\,\frac{1-\left(\frac{x}{2}\right)^k}{1-\frac{x}{2}}\qquad\qquad\qquad.
\end{multline*}
Denoting $x=2t,\,\, a^2=\tau$, we rewrite this formula as
\begin{equation}\label{10}
N_k(t;\tau)=\frac{1}{\tau(1-t)}\left(\tau +(1-\tau)t^k\right),\qquad (k\geq1,\, 0<t<1,\, \tau>0).
\end{equation}
Substituting \reff{10} in the formula \reff{03} for the Mandel parameter (and taking into account that
$\ds\frac{d}{dx}=\frac{1}{2}\,\frac{d}{dt}$) we get
\begin{equation}\label{13}
Q_k(t;\tau)=t\left(\frac{(N_k(t;\tau))_{t^2}^{\prime\prime}}{(N_k(t;\tau))_{t}^{\prime}}-
\frac{(N_k(t;\tau))_{t}^{\prime}}{N_k(t;\tau)}\right).
\end{equation}

For the sake of brevity, we will not specify the dependency on $\tau$ in the future. Let us note that
\begin{equation*}
N_k(t)\,N_k^{\prime}(t)=\frac{1}{2} \left(N_k^{2}(t)\right)_t^{\prime}
\end{equation*}
and denote
\begin{equation}\label{14}
q_k(t)=N_k^{\prime\prime}(t)N_k(t)-\left(N_k^{\prime}(t)\right)^2.
\end{equation}
Then \reff{13} can be rewritten in the form
\begin{equation}\label{15}
Q_k(t)=\frac{2t}{(N_k^2)^\prime_t}\,q_k(t).
\end{equation}

Recall that we are interested in the sign of $Q_k (t)$ for $0<t<1,\, \tau>0$. We show that this sign
coincides with the sign of $q_k(t)$. To do this, it is sufficient to prove that $(N_k^2)^\prime_t>0$.
We show that the function $N_k(t;\tau)$ increases monotonically on the interval $0<t<1$ for all $\tau>0$.
Indeed, the formula \reff{10} can be rewritten as
\begin{equation}\label{16}
N_k(t;\tau)=\frac{1-t^k}{1-t}+\frac{1}{\tau}\,\frac{t^k}{1-t}=
(1+t+\ldots+t^{k-1})+\frac{1}{\tau}\,\frac{t^k}{1-t}.
\end{equation}
The first term on the right side of \reff{16} increases monotonically with grow of $t$, and in the
second term, for any $\tau>0$, the numerator monotonically increases, and the denominator
monotonically decreases with increasing $t$ in the interval $(0;1)$. Therefore, the function
$N_k(t;\tau)$, and hence $N_k^2(t;\tau)$, increases monotonically over the interval $(0;1)$
for any $\tau>0$. It follows that $(N_k^2)^\prime_t>0$. So we have shown that
\begin{equation}
\text{sign}(Q_k(t;\tau))=\text{sign}(q_k(t;\tau)), \qquad 0<t<1,\, \tau>0.
\end{equation}

Thus, we need to calculate $q_k(t;\tau)$. From the formula \reff{10} we have
\begin{equation}
(N_k)^{\prime}_t(t)=\frac{1}{\tau(1-t)^2}\left(\tau+(1-\tau)(kt^{k-1}+(1-k)t^k)\right),
\end{equation}
\begin{multline}\label{18}
(N_k)^{\prime\prime}_t(t)=\frac{1}{\tau(1-t)^3}\times\\
\left(2\tau+(1-\tau)(k(k-1)-2k(k-2)t+(k-1)(k-2)t^2)t^{k-2}\right).
\end{multline}
Then from \reff{14}-\reff{18} follows
\begin{equation}\label{19}
q_k(t;\tau)=\frac{1}{\tau^2(1-t)^4}\,P_k(t;\tau)
\end{equation}
where
\begin{multline}\label{20}
P_k(t;\tau)=\left[2\tau+(1-\tau)t^{k-2}\left((k-1)(k-2)t^2-2k(k-2)t+k(k-1)\right)\right]\times \\
(\tau+(1-\tau)t^k)-\left(\tau+(1-\tau)t^{k-1}(k+(1-k)t)\right)^2.
\end{multline}
It follows from \reff{19} that $\text{sign}(q_k(t;\tau))=\text{sign}(P_k(t;\tau))$
when $0<t<1,\,\, \tau>0$. Therefore, the problem was reduced to determining the sign of the polynomial
$P_k(t;\tau)$, which is convenient to write as
\begin{multline}
P_k(t;\tau)=\tau^2+\tau(1-\tau)k(k-1)t^{k-2}-2\tau(1-\tau) k(k-1)t^{k-1}+\\
\tau(1-\tau)(k(k-1)+2)t^k-(1-\tau)^2kt^{2k-2}+\\
2(1-\tau)^2kt^{2k-1}-(1-\tau)^2(k-1)t^{2k}.
\end{multline}

So, to determine the sign of the Mandel parameter $Q_M(t;\tau)$ in the band $0<t<1,\,\, \tau>0,$ 
we need to find the roots and areas of the constant sign for the polynomial $P_k(t;\tau)$ in this band .
We do not consider the solution of this general problem in this paper, but we will discuss the
technical difficulties that arise using the example of $P_2(t;\tau)$.

\section{Investigation of the sign of the Mandel \\ parameter for $k=2$}
For $k=2$, the formula \reff{20} takes the form
\begin{equation}
P_2(t;\tau)=\tau(2-\tau)+4\tau(\tau-1)t+2(1-\tau)(3\tau-1)t^2+4(\tau-1)^2t^3-(\tau-1)^2t^4.
\end{equation}
Our task is to investigate $\text{sign}(P_2(t;\tau))$ in the region $\Pi:0<t<1,\,\, \tau>0$. We divide
the $\Pi$ region into three parts
\begin{equation*}
\Pi=\Pi_1\,\cup\,\Pi_2\,\cup\,\Pi_3,
\end{equation*}
where
\begin{eqnarray}
  \Pi_1&=&\{0<t<1,\,\, 0<\tau\leq 1\} \nn \\
  \Pi_2&=&\{0<t<1,\,\, 1<\tau\leq 2\} \\
  \Pi_3&=&\{0<t<1,\,\, \tau>2\}. \nn
\end{eqnarray}

Let us first consider the case $\Pi_2$ (which also includes the classical Chebyshev polynomials of the 1st and 2nd kind).
We prove that in the region $\Pi_2$ the Mandel parameter is positive, i.e.
\begin{equation*}
\text{sign}(P_2(t;\tau))>0\quad\text{}\quad 0<t<1,\, 1<\tau\leq 2.
\end{equation*}
To prove this, we divide the terms on the right hand side of the equation
into the following parts:
\begin{eqnarray*}
  \sigma_1&=&\tau(2-\tau)=1-(\tau-1)^2; \\
  \sigma_2&=&4\tau(\tau-1)t; \\
  \sigma_3&=&2(1-\tau)(3\tau-1)t^2=-2(\tau-1)^2t^2+4\tau(1-\tau)t^2=\sigma_{31}+\sigma_{32}; \\
  \sigma_4&=&4(\tau-1)^2t^3=3(\tau-1)^2t^3+(\tau-1)^2t^3=\sigma_{41}+\sigma_{42}; \\
  \sigma_5&=&-(\tau-1)^2t^4.
\end{eqnarray*}
Next, we prove that the following inequalities hold in the region $\Pi_2$:
\begin{eqnarray*}
\gamma_1&=&(\sigma_1+\sigma_{31}+\sigma_{41})>0; \\
\gamma_2&=&(\sigma_2+\sigma_{32})>0; \\
\gamma_3&=&(\sigma_{42}+\sigma_5)>0.
\end{eqnarray*}

The validity of the inequalities $\gamma_2$ and $\gamma_3$ is quite obvious. Really,
\begin{equation*}
\gamma_2=4\tau(\tau-1)t(1-t)>0\quad\text{and}\quad \gamma_3=(\tau-1)^2(1-t)t^3>0,
\end{equation*}
since in the domain under consideration, all factors are positive.

Consider now $\gamma_1=\gamma_1(t;\tau)$. We have
\begin{equation*}
\gamma_1(t;\tau)=1-(\tau-1)^2-2(\tau-1)^2t^2+3(\tau-1)^2t^3.
\end{equation*}
Let's denote $\xi=\tau-1$. Then $0<\xi<1$,
\begin{equation}\label{26}
\gamma_1(t;\tau)=\gamma_1(t;\xi)=1-\xi^2(3t^3-2t^2-1).
\end{equation}

To prove that $\gamma_1(t;\tau)$ is positive, it is
sufficient to show that the function $\psi(t)=3t^3-2t^2-1$ satisfies the inequality $\psi(t)\leq1$ on the
interval $0<t<1$. The latter inequality holds, since a simple analysis on the extremum
shows that in fact, even the stronger inequality $\psi(t)\leq 0$ is true on this interval.

So we proved that in the domain $\Pi_2$, the Mandel parameter is positive.
\smallskip

Unfortunately, in the regions $\Pi_1$ and $\Pi_3$, the situation is much more complicated ---
the boundaries of the sections in which
the Mandel parameter has a constant sign can be set for fixed values of the parameter $a$ only approximately,
using numerical methods. The tables below allows to determine the nature of the location
of the sign-constant regions of the Mandel parameter.
In these tables, we return to the original variables $x$ and $a$ in which
the Mandel parrameter for $k=2$ has the form
\begin{multline*}
\qquad Q_M(x;2;a)=\\
-\frac{x(a^4(x-2)^4+x^2(8-8 x+x^2)-2a^2(16-16x+16x^2-8x^3+x^4))}{(x-2)(a^2(x-2)^2-(x-4)x)(a^2(x^2-4)-x^2)}.
\end{multline*}

For the area $\Pi_3$ with $a>\sqrt{2}$, we have

\begin{center}
\begin{tabular}{|c|c|c|c|}
\hline
$a$&$Q_M>0$&$Q_M<0$&$Q_M(x_{min})$\\
\hline
$1,5$&$x>0,108$&$x\in(0;0,107)$&$Q^{min}(0,054)=-0,0015$\\
\hline
$2,0$&$x>0,47$&$x\in(0;0,46)$&$Q^{min}(0,247)=-0,0325$\\
\hline
$2,5$&$x>0,66$&$x\in(0;0,65)$&$Q^{min}(0,363)=-0,0666$\\
\hline
$3,0$&$x>0,79$&$x\in(0;0,78)$&$Q^{min}(0,450)=-0,0961$\\
\hline
$4,0$&$x>0,97$&$x\in(0;0,96)$&$Q^{min}(0,581)=-0,1419$\\
\hline
$5,0$&$x>1,09$&$x\in(0;1,08)$&$Q^{min}(0,678)=-0,1755$\\
\hline
$6,0$&$x>1,18$&$x\in(0;1,17)$&$Q^{min}(0,755)=-0,2015$\\
\hline
$7,0$&$x>1,25$&$x\in(0;1,24)$&$Q^{min}(0,818)=-0,2222$\\
\hline
$8,0$&$x>1,31$&$x\in(0;1,30)$&$Q^{min}(0,871)=-0,2392$\\
\hline
$9,0$&$x>1,36$&$x\in(0;1,35)$&$Q^{min}(0,918)=-0,2534$\\
\hline
$10$&$x>1,39$&$x\in(0;1,38)$&$Q^{min}(0,958)=-0,2656$\\
\hline
$20$&$x>1,60$&$x\in(0;1,59)$&$Q^{min}(1,2005)=-0,3331$\\
\hline
$100$&$x>1,86$&$x\in(0;1,85)$&$Q^{min}(1,5989)=-0,4255$\\
\hline
\end{tabular}
\end{center}

From the results presented in this table, it follows that when the parameter $a$
increases, the interval in which $Q_M$ is negative increases, the minimum value of $Q^{min}$
is shifted to the right end of the interval, and its absolute value $|Q^{min}|$ increases.

In the band $\Pi_1:0<a<1$, the situation is somewhat different, as can be seen from the following table.
\begin{center}
\begin{tabular}{|c|c|c|c|}
\hline
$a$&$Q_M>0$&$Q_M<0$&$Q_M(x_{min})$\\
\hline
$0,01$&$x\in(0;0,020)$&$x\in(0,021;1,17)$&$Q^{min}(0,168)=-0,925$\\
&$x\in(1,18;2)$&&\\
\hline
$0,1$&$x\in(0;0,20)$&$x\in(0,21;1,14)$&$Q^{min}(0,575)=-0,563$\\
&$x\in(1,15;2)$&&\\
\hline
$0,2$&$x\in(0;0,43)$&$x\in(0,44;1,106)$&$Q^{min}(0,742)=-0,223$\\
&$x\in(1,07;2)$&&\\
\hline
$0,25$&$x\in(0;0,61)$&$x\in(0,62;0,96)$&$Q^{min}(0,787)=-0,066$\\
&$x\in(0,97;2)$&&\\
\hline
$0,3$&$x\in(0;2)$&---&$Q^{min}(0,812)=0,080$\\
\hline
\end{tabular}
\end{center}

As $a$ increases, the interval in which $Q_M$ is negative
and the absolute value $|Q^{min}|$ decreases. For $0.3<a<1$, the Mandel parameter is positive.
\bigskip

{\bf Acknowledgements} Authors are grateful to I. K. Litskevich for assistance in performing some calculations.

\end{document}